\documentstyle[twoside,fleqn,espcrc2,psfig]{article}

\newcommand{\AmS}{{\protect\the\textfont2
  A\kern-.1667em\lower.5ex\hbox{M}\kern-.125emS}}

\def\I{{\cal I}}
\def\eps{\epsilon}
\def\e{\epsilon}
\def\Ord{{\cal O}}
\def\M{{\cal M}}
\def\tr{\mathop{\rm tr}\nolimits}
\def\Tr{\mathop{\rm Tr}\nolimits}
\def\P{{\rm P}}
\def\NP{{\rm NP}}
\def\pol{\varepsilon}
\def\Ksl{\s{K}}
\def\ksl{\s{k}}

\newbox\charbox
\newbox\slabox
\def\s#1{{      % Feynman slash
        \setbox\charbox=\hbox{$#1$}
        \setbox\slabox=\hbox{$/$}
        \dimen\charbox=\ht\slabox
        \advance\dimen\charbox by -\dp\slabox
        \advance\dimen\charbox by -\ht\charbox
        \advance\dimen\charbox by \dp\charbox
        \divide\dimen\charbox by 2
        \raise-\dimen\charbox\hbox to \wd\charbox{\hss/\hss}
        \llap{$#1$}
}}

\def\spa#1.#2{\left\langle#1\,#2\right\rangle}
\def\spb#1.#2{\left[#1\,#2\right]}
\def\spab#1.#2.#3{\langle\mskip-1mu{#1}^- 
                  | #2 | {#3}^-\mskip-1mu\rangle}
\def\spba#1.#2.#3{\langle\mskip-1mu{#1}^+ 
                  | #2 | {#3}^+\mskip-1mu\rangle}
\def\spbb#1.#2.#3{\langle\mskip-1mu{#1}^+ 
                  | #2 | {#3}^-\mskip-1mu\rangle}
\def\lor#1.#2{\left(#1\,#2\right)}

\def\tree{{\rm tree}}
\def\oneloop{{1 \mbox{-} \rm loop}}
\def\twoloop{{2 \mbox{-} \rm loop}}

\def\eqn#1{eq.~(\ref{#1})}

\def\fig#1{fig.~{\ref{#1}}}

\newskip\humongous \humongous=0pt plus 1000pt minus 1000pt
\def\caja{\mathsurround=0pt}
\def\eqalign#1{\,\vcenter{\openup1\jot \caja
        \ialign{\strut \hfil$\displaystyle{##}$&$
        \displaystyle{{}##}$\hfil\crcr#1\crcr}}\,}
\newif\ifdtup

\newcounter{eqnumber}
\renewcommand{\theeqnumber}{\arabic{eqnumber}}
\def\equn{
\refstepcounter{eqnumber}
\eqno({\rm \theeqnumber})
}

\begin{document}

% declarations for front matter

\title{
\vskip -1. cm  {\normalsize hep-th/0002078\hfill SLAC-PUB-8348\\
                                   \hfill UCLA/00/TEP/4 \\ 
                                   \hfill HUTP-00/A002  \\ 
                                   \hfill SWAT-00-250   \\ 
                                   \hfill UFIFT-HEP-00-01\\$\null$}
On Perturbative Gravity and Gauge Theory\thanks{This research 
was supported by the US Department of Energy under grants DE-FG03-91ER40662, 
DE-AC03-76SF00515 and DE-FG02-97ER41029.}  }

\author{
Z. Bern,\address{Department of Physics, UCLA, Los Angeles, CA 90095}%
\thanks{Presenter
at Third Meeting on 
Constrained Dynamics and Quantum Gravity, Villasimius (Sardinia, Italy)
September 13-17, 1999 and at the 
Workshop on Light-Cone QCD and Nonperturbative Hadron Physics, 
University of Adelaide (Australia) December 13-22, 1999.}
L. Dixon,\address{Stanford Linear Accelerator Center, Stanford University,
                      Stanford, CA 94309}
D.C. Dunbar,\address{Department of Physics, University of Wales Swansea,
            Swansea SA2 8PP, UK}
A.K. Grant,\address{Department of Physics, Harvard University,
     Cambridge, MA 02138}
M. Perelstein,\hbox{$^{\rm b}$}
and 
J.S. Rozowsky\address{Institute for Fundamental Theory,  
Department of Physics, University of Florida, Gainesville, FL 32611} }
       
\begin{abstract}
We review some applications of tree-level (classical) relations
between gravity and gauge theory that follow from string theory.
Together with $D$-dimensional unitarity, these
relations can be used to perturbatively quantize gravity theories,
i.e. they contain the necessary information for obtaining loop
contributions.  We also review recent applications of these ideas
showing that $N=1$ $D=11$ supergravity diverges, and review
arguments that $N=8$ $D=4$ supergravity is less divergent than
previously thought, though it does appear to diverge at five loops.
Finally, we describe field variables for the Einstein-Hilbert
Lagrangian that help clarify the perturbative relationship between
gravity and gauge theory.
\end{abstract}

\maketitle

%%%%%%%%%%%%%%%%%%%%%%%%%%%%%%%%%%%%%%%%%%%%%%%%%%%%%%%%%%%%%%%%%%%%
\section{Introduction}

In this talk we describe recent
work~\cite{BDDPR,AllPlusGrav,MHVGrav,Aaron,Square} on perturbative
relations between gravity and gauge theories.  Although gauge and
gravity theories are similar in that they both have a local symmetry,
their dynamical behavior is quite different. For example, in four
dimensions gauge theories are renormalizable and exhibit asymptotic
freedom, neither of which property is shared by gravity.  The
structures of the Lagrangians are also rather different: the Yang-Mills
Lagrangian contains only up to four-point interactions while the
Einstein-Hilbert Lagrangian contains infinitely many interactions.
Nevertheless, in the context of perturbation theory, it turns out that
tree-level gravity amplitudes can, roughly speaking, be expressed as a
sum of `squares' of gauge theory amplitudes.  These tree-level
(classical) relations between gravity and gauge theory amplitudes
follow from the Kawai, Lewellen and Tye (KLT)~\cite{KLT} relations
between open and closed string tree amplitudes.  When combined with
the $D$-dimensional unitarity methods described in
refs.~\cite{SusyFour,Review}, it provides a useful tool for
investigating the ultra-violet behavior of gravity field theories.
The unitarity methods have also been applied to QCD loop computations
of phenomenological interest and to supersymmetric gauge theory
computations~\cite{SusyFour,BRY}.

Ultraviolet properties are one of the central issues for perturbative
gravity.  Although gravity is non-renormalizable by power counting, no
divergence has, in fact, been established by a direct calculation for
any supersymmetric theory of gravity in four dimensions.  Explicit
calculations have established that non-supersymmetric theories of
gravity with matter generically diverge at one
loop~\cite{tHooftVeltmanAnnPoin,tHooftGrav,DeserEtal}, and pure
gravity diverges at two loops~\cite{PureGravityInfinityGSV}.  However,
in any supergravity theory in $D=4$, supersymmetry Ward
identities~\cite{SWI} forbid all possible one-loop~\cite{OneLoopSUGRA}
and two-loop~\cite{Grisaru} counterterms.  Thus, at least a three-loop
calculation is required to definitively address the question of
divergences in four-dimensional supergravity.  There is a candidate
counterterm at three loops for all supergravities including the
maximally extended version ($N=8$)~\cite{KalloshNeight,HoweStelle}.
However, no explicit three loop (super) gravity calculations have
appeared.  It is in principle possible that the coefficient of a
potential counterterm can vanish, especially if the full symmetry of
the theory is taken into account.  In ref.~\cite{BDDPR} it was argued
that this may indeed be the case for the potential three-loop counterterm
of $N=8$ $D=4$ supergravity.  For there to be cancellations of
divergences beyond this, the theory would require a symmetry which we
do not fully understand yet.

\begin{figure}[t]
\centerline{\psfig{figure=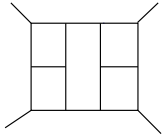,height=.7in}}
\vskip -.4cm 
\caption{An example of a five-loop diagram.}
\label{fig:Multiloop}
\end{figure}

With traditional perturbative approaches~\cite{DeWitt} to performing
explicit calculations, as the number of loops increases the number of
algebraic terms proliferates rapidly beyond the point where
computations are practical.  Consider the five-loop diagram in
\fig{fig:Multiloop} (which, as described below, is of interest for
ultraviolet divergences in $N=8$ $D=4$ supergravity).  In de Donder
gauge the pure graviton diagram contains twelve vertices, each of the
order of a hundred terms, and sixteen graviton propagators, each with
three terms.  When expanded, this yields a total of roughly $10^{30}$
terms, even before performing any loop integrations.  Needless to say,
this is well beyond what can be reasonably implemented on any
computer.

%%%%%%%%%
%%%%%%%

Our approach for dealing with this is to reformulate the quantization
of gravity in the context of perturbation theory using the KLT
relations together with $D$-dimensional unitarity.  

The final topic of this talk will cover some recent progress in
understanding the KLT relations from a Lagrangian point of
view~\cite{Aaron}.  It turns out that with an appropriate choice of
field variables one can separate the Lorentz indices appearing in the
Lagrangian into `left' and `right' classes~\cite{Siegel,Aaron},
mimicking the similar separation that occurs in string theory.
Moreover, with further field redefinitions and a non-linear gauge
choice it is possible to arrange the off-shell three graviton vertex
so that it is expressible in terms of a sum of squares of Yang-Mills
three gluon vertices.

\section{Method for Investigating Perturbative Gravity}

Our formulation of perturbative quantum gravity is based on two ingredients:
\begin{enumerate}
\item The Kawai, Lewellen and Tye relations between closed
and open string tree-level S-matrices~\cite{KLT}. 
\item The observation that the $D$-dimensional tree amplitudes contain
all information necessary for building the complete perturbative
$S$-matrix to all loop orders~\cite{SusyFour,Review}.
\end{enumerate}

\subsection{The KLT tree-level relations.}

In the field theory limit ($\alpha' \to 0$) the KLT relations for the four-
and five-point amplitudes are~\cite{BGK}
$$
\eqalign{
& M_4^{\rm tree}  (1,2,3,4) = \cr
& \hskip .3 cm 
 - i s_{12} A_4^{\rm tree} (1,2,3,4) \, A_4^{\rm tree}(1,2,4,3)\,, \cr
& M_5^{\rm tree}(1,2,3,4,5) = \cr
& \hskip .3 cm 
i s_{12} s_{34}  A_5^{\rm tree}(1,2,3,4,5)
                                     A_5^{\rm tree}(2,1,4,3,5) \cr
&
 + i s_{13}s_{24} A_5^{\rm tree}(1,3,2,4,5) \, A_5^{\rm tree}(3,1,4,2,5)\,,\cr}
\equn\label{KLTExamples}
$$
where the $M_n$'s are the amplitudes in a gravity theory stripped of
couplings, the $A_n$'s are the color-ordered sub-amplitudes in a gauge
theory and $s_{ij}\equiv (k_i+k_j)^2$.  We suppress all $\pol_j$
polarizations and $k_j$ momenta, but keep the `$j$' labels to
distinguish the external legs. Full gauge theory amplitudes 
are given in terms of the partial amplitudes $A_n$, via
$$
\eqalign{
{\cal A}_n^{\rm tree}& (1,2,\ldots n) = 
 g^{(n-2)} \sum_{\sigma \in S_n/Z_n} \cr
&
{\rm Tr}\left( T^{a_{\sigma(1)}} 
\cdots  T^{a_{\sigma(n)}} \right)
 A_n^{\rm tree}(\sigma(1), \ldots, \sigma(n)) \,, \cr} \hskip .6 cm 
$$
where $S_n/Z_n$ is the set of all permutations, but with cyclic
rotations removed, and $g$ is the gauge theory coupling constant.
The $T^{a_i}$ are fundamental representation
matrices for the Yang-Mills gauge group $SU(N_c)$, normalized so that
$\Tr(T^aT^b) = \delta^{ab}$.  
For states coupling with the strength of gravity, 
the full amplitudes including the gravitational coupling constant are, 
$$
{\cal M}_n^{\rm tree} (1,\ldots n) = 
\left({  \kappa \over 2} \right)^{(n-2)} 
M_n^{\rm tree}(1,\ldots n)\,, \hskip 1 cm 
$$
where $\kappa^2 = 32\pi G_N$.
The KLT equations generically hold for any closed string states, using
their Fock space factorization into pairs of open string states.

Berends, Giele and Kuijf~\cite{BGK} exploited the KLT
relations~(\ref{KLTExamples}) and their $n$-point generalizations to
obtain an infinite set of maximally helicity violating (MHV) graviton
tree amplitudes, using the known MHV Yang-Mills
amplitudes~\cite{ParkeTaylor}.  Cases of gauge theory coupled to
gravity have very recently been discussed in ref.~\cite{Square}.
Interestingly, the color charges associated with any gauge fields
appearing in gravity theories are represented through the KLT
equations as flavor charges carried either by scalars or fermions.
For example, by applying the KLT equations the three-gluon
one-graviton amplitude may be expressed as
$$
\eqalign{
& {\cal M}_4^\tree  (1^-_g, 2^-_g, 3^+_g, 4^+_h)
  =-i g {\kappa \over 2} 
s_{12} \cr
& \hskip .2 cm \times
A_4^\tree(1^-_g, 2^-_g, 3^+_g, 4^+_g) \times
       A_4^\tree(1_s, 2_s, 4^+_g, 3_s) \hskip 2 cm \cr
& \hskip .2 cm 
= g  {\kappa \over 2}  {\spa1.2^4 \over \spa1.2\spa2.3\spa3.4\spa4.1} 
\times \sqrt{2} f^{a_1 a_2a_3} \, {\spb4.3\spa3.2 \over \spa2.4} \,,\cr}
$$
where the $\pm$ superscripts denote the helicities and the subscripts
$h$, $g$ and $s$ denote whether a given leg is a graviton, gluon or
scalar.  On the right-hand side of the equation, the group theory
indices are flavor indices for the scalars.  On the left-hand side
they are reinterpreted as color indices for gluons. For simplicity,
the amplitudes have been expressed in terms of $D=4$ spinor inner
products (see e.g. ref.~\cite{ManganoReview}), although the
factorization of the amplitude into purely gauge theory amplitudes
holds in any dimension.  The spinor inner products are denoted by
$\spa{i}.j = \langle i^- | j^+\rangle$ and $\spb{i}.j = \langle i^+|
j^-\rangle$, where $|i^{\pm}\rangle$ are massless Weyl spinors of
momentum $k_i$, labeled with the sign of the helicity.  They are
antisymmetric, with norm $|\spa{i}.j| = |\spb{i}.j| = \sqrt{s_{ij}}$.

%%%%%%%%%%%%%%%%%%%%%

\subsection{Cut Construction of Loop Amplitudes}

We now outline the use of the KLT relations for computing multi-loop
gravity amplitudes, starting from gauge theory amplitudes. Although the KLT
equations hold only at the classical tree-level,
$D$-dimensional unitarity considerations can be used to extend them to
the quantum level.  The application of $D$-dimensional unitarity has
been extensively discussed for the case of gauge theory amplitudes
\cite{SusyFour,Review} so here we describe it only briefly.

\begin{figure}[t]
\centerline{\psfig{figure=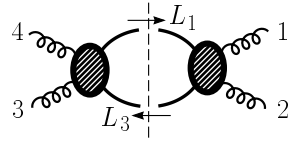,height=.8in}}
\vskip -.4cm 
\caption{The two-particle cut at one loop in the channel carrying 
momentum $k_1+k_2$.}
\label{fig:TwoParticle}
\end{figure}

The unitarity cuts of a loop amplitude can be expressed in terms of
amplitudes containing fewer loops.  For example, the two-particle 
cut of a one-loop
four-point amplitude in the channel carrying momentum $k_1 + k_2$, as
shown in \fig{fig:TwoParticle}, can be expressed as the cut of,
$$
\eqalign{
 \sum_{\rm states} & \int {d^D L_1 \over (4 \pi)^D} \, 
  {i\over L_1^2} \,
{\cal M}_{4}^\tree(-L_1,1,2,L_3) \cr
& \hskip 1. cm 
\times
\,{i \over L_3^2}\,
{\cal M}_{4}^\tree (-L_3,3,4,L_1) \Bigr|_{\rm cut} \,, 
\cr}
\equn\label{BasicCutEquation}
$$
where $L_3 = L_1 - k_1 - k_2$, and the sum runs over all states
crossing the cut. We label $D$-dimensional momenta with capital
letters and four-dimensional ones with lower case.  We apply the
on-shell conditions $L_1^2 = L_3^2 = 0$ to the amplitudes appearing in
the cut even though the loop momentum is unrestricted; only functions
with a cut in the given channel under consideration are reliably
computed in this way.

% FIGURE
%%%%%%%%%
%%%%%%%

Complete amplitudes are found by combining all cuts into a single
function with the correct cuts in all channels.  If one works with an
arbitrary dimension $D$ in \eqn{BasicCutEquation}, and takes care to
keep the full analytic behavior as a function of $D$, then the results
will be free of subtraction ambiguities that are commonly present in
cutting methods~\cite{ExactUnitarity,SusyFour,Review}. (The regularization
scheme dependence remains, of course.)

An important advantage of the cutting approach is that the
gauge-invariant amplitudes on either side of the cut may be simplified
before attempting to evaluate the cut integral~\cite{Review}.

%%%%%%%%%%%%%%%%%%%%%%%%%%%%%%%%%%%%%
\section{Recycling Gauge Theory Into Gravity Loop amplitudes}

Consider the one-loop amplitude with 
four identical helicity gravitons and a scalar 
in the loop~\cite{AllPlusGrav,MHVGrav}. The cut in the $s_{12}$ channel is
$$
\eqalign{
\int {d^D L_1 \over (2\pi)^D} \; & {i \over L_1^2} 
 M_4^\tree(-L_1^s, 1^+, 2^+, L_3^s) \, {i \over L_3^2} \cr
&  \hskip 1 cm \times 
M_4^\tree(-L_3^s, 3^+, 4^+, L_1^s)\Bigr|_{\rm cut} \,, } \hskip 1 cm 
$$
where the superscript $s$ indicates that the cut lines are scalars.
Using the KLT expressions (\ref{KLTExamples}) we may replace the
gravity tree amplitudes appearing in the cuts with products of gauge
theory amplitudes.  The required gauge theory tree amplitudes, with
two external scalar legs and two gluons, are relatively simple to
obtain using Feynman diagrams and are,
$$
\eqalign{
& A_4^\tree(-L_1^s,1^+,2^+,L_3^s) =  i
{\mu^2\spb1.2\over\spa1.2 [( \ell_1 -k_1)^2 -\mu^2] }\,, \cr
& A_4^\tree(-L_1^s, 1^+, L_3^s, 2^+) 
 = - i {\mu^2\spb1.2\over\spa1.2} \cr
& \hskip .5 cm  \times 
 \biggl[{1\over (\ell_1 -k_1)^2 -\mu^2}
 + {1\over (\ell_1 -k_2)^2 -\mu^2}\biggr] \,, \cr}
$$
where $L_1 = \ell_1 + \mu$.
The gluon momenta are four-dimensional, but the scalar momenta
are allowed to have a $(-2\e)$-dimensional component $\vec{\mu}$, with
$\vec{\mu}\cdot\vec{\mu} = \mu^2 > 0$.  The overall factor of $\mu^2$
appearing in these tree amplitudes means that they vanish in the
four-dimensional limit, in accord with a supersymmetry Ward
identity~\cite{SWI}.  In the KLT relation (\ref{KLTExamples}), one of
the propagators cancels, leaving
$$
\eqalign{
& M_4^\tree(-L_1^s, 1^+, 2^+, L_3^s) = 
- i \biggl({\mu^2\spb1.2\over\spa1.2}\biggr)^2 \cr
& \hskip 1 cm \times
 \biggl[{1\over (\ell_1 -k_1)^2-\mu^2}
 + {1\over (\ell_1 -k_2)^2-\mu^2}\biggr] \,. }
$$
By symmetry, the tree amplitudes appearing in any of the other
cuts are the same up to relabelings.

\begin{figure}[t]
\centerline{\psfig{figure=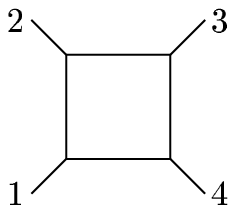,height=.7in}}
\vskip -.4cm 
\caption{The one loop box integral. }
\label{OneLoopIntegralFigure}
\end{figure}

Combining all three cuts into a single function that has the correct
cuts in all channels yields
$$
\eqalign{
& M_4^\oneloop(1^+, 2^+, 3^+, 4^+)  = 
    2 {\spb1.2^2\spb3.4^2\over\spa1.2^2\spa3.4^2} \cr
& \hskip 1 cm \times
\Bigl(\I_4^\oneloop[\mu^8](s,t) + 
      \I_4^\oneloop[\mu^8](s,u) \cr
& \hskip 3 cm 
   + \I_4^\oneloop[\mu^8](t,u) \Bigr)\,, \cr } \hskip 2. cm 
\equn\label{FourGravAllPlus}
$$
where $s = s_{12}, \; t = s_{14}, \; u= s_{13}$ are the usual Mandelstam 
variables and
$$
\eqalign{
& \I_4^\oneloop[{\cal P}](s,t) = 
\int {d^D L \over (2\pi)^D} 
 \cr
& \hskip .5 cm  \times
\, {{\cal P} \over L^2 (L - k_1)^2
       (L - k_1 - k_2)^2
        (L + k_4)^2  } \cr}
\equn\label{OneLoopIntegral} 
$$
is the scalar box integral depicted in \fig{OneLoopIntegralFigure}
 with the external legs arranged in the 
order 1234. In \eqn{FourGravAllPlus} the numerator $\cal P$ is
$\mu^8$.  The two other scalar integrals that appear correspond 
to the two other distinct orderings of the four external legs.
The spinor factor $\spb1.2^2\spb3.4^2/(\spa1.2^2\spa3.4^2)$ in
\eqn{FourGravAllPlus} is actually completely symmetric, although not
manifestly so.  By rewriting this factor and extracting the leading
$\Ord(\eps^0)$ contribution from the integral, the final one-loop
$D=4$ result after reinserting the gravitational coupling is
$$
\eqalign{
{\cal M}_4^\oneloop &(1^+, 2^+, 3^+, 4^+)  = - \, {i \over (4 \pi)^2} 
\Bigl({\kappa\over 2} \Bigr)^4\, \cr
& \times 
\biggl( {st \over \spa1.2 \spa2.3 \spa3.4 \spa4.1} \biggr)^2 
{s^2 + t^2 + u^2 \over 120} \,, \cr}
\equn\label{FourGravPlus}
$$
in agreement with a previous calculation~\cite{DN}.

% FIGURE
%%%%%%%%%
%%%%%%%

This can be extended to an arbitrary number of external legs.  Using
the cutting methods we have also calculated the five- and six-point
amplitudes.  By demanding that the amplitudes have the correct
factorization properties as momenta become either soft or
collinear~\cite{SusyFour,Factorization} we have found an ansatz for
the one-loop maximally helicity violating amplitudes for an arbitrary
number of external legs\cite{AllPlusGrav,MHVGrav},
$$
\eqalign{
& \M_n^\oneloop(1^+, 2^+, \ldots, n^+) = 
- {i \over (4 \pi)^2\cdot 960} \, \left({-\kappa\over 2}\right)^n \cr
& \hskip .4 cm \times \hskip -.3 cm 
\sum_{1 \leq a < b \leq n \atop M, N} h(a, M, b) h(b, N, a) 
 \tr^3[a\, M\, b\, N]\,, \cr}
\equn\label{AllNPlus}
$$
where $a$ and $b$ are massless legs, and $M$ and $N$ are two sets forming 
a distinct nontrivial partition of the remaining $n-2$ legs.
Also, $\tr[a\, M\, b\, N] \equiv \tr[\ksl_a \Ksl_{\! M} \ksl_b \Ksl_{\! N}]$,
where $K_M$ is the sum of momenta of the legs in the set $M$.
The rational function $h$ is a bit more complicated and
may be found in refs.~\cite{AllPlusGrav,MHVGrav}.

These amplitudes are not ultraviolet divergent and thus do not depend
on a cutoff. By unitarity considerations any fundamental theory of
gravity must therefore necessarily yield the one-loop amplitudes
(\ref{FourGravPlus}) and (\ref{AllNPlus}) in the low energy limit.
(The tree amplitude can, however, have a contribution of the same
dimension as that of the one-loop amplitude; this contribution would
depend on the details of the underlying fundamental theory.)  The
ability to obtain exact expressions for gravity loop amplitudes
demonstrates the utility of this approach for investigating quantum
properties of gravity theories.  As our next example, we turn to the
divergence properties of maximally supersymmetric theories.

%%%%%%%%%%%%%%%%%%%%%%%%%%%%%%%%%%%%%%%%%%%%%%%%%%%%%%%%%
\section{Maximal Supergravity}

Maximal $N=8$ supergravity can be expected to be the least divergent of the
four-dimensional supergravity theories due to its high degree of
symmetry.  Moreover, from a technical viewpoint maximally
supersymmetric $N=8$ amplitudes are by far the easiest to deal with in
our formalism because of spectacular supersymmetric cancellations.
For these reasons it is logical to re-investigate the divergence
properties of this theory first~\cite{BDDPR}.  It
should be possible to apply similar methods to theories with less
supersymmetry.

\subsection{Cut Construction}

Again we obtain supergravity amplitudes by recycling gauge theory
calculations.  For the $N=8$ case, we factorize each of the 256 states
of the multiplet into a tensor product of $N=4$ super-Yang-Mills
states.  The key equation for obtaining the two-particle cuts is,
$$
\eqalign{
& \sum_{N=8 \atop \rm\ states}  
M_4^\tree(-L_1,  1, 2, L_3) \times
  M_4^\tree(-L_3, 3, 4, L_1) \cr
& =
s^2  \!\sum_{N=4 \atop \rm\ states} \!
  A_4^\tree(-L_1,  1, 2, L_3) \times
     A_4^\tree(-L_3,  3,4 , L_1)
     \cr
\null & \hskip .4 truecm
\times
\!\sum_{N=4 \atop \rm\ states} \!
 A_4^\tree(L_3, 1, 2 , -L_1) \times
                A_4^\tree(L_1, 3, 4, -L_3) \,,\cr}
\equn\label{GravitySewingStart}
$$
where the sum on the left-hand side runs over all states in the $N=8$
supergravity multiplet. On the right-hand side the two sums run over
the states of the $N=4$ super-Yang-Mills multiplet: a gluon, four Weyl
fermions and six real scalars.  Given the corresponding $N=4$
Yang-Mills two-particle sewing equation~\cite{BRY},
$$
\eqalign{
& \sum_{N=4\atop \rm  states}
 A_4^\tree(-L_1, 1, 2, L_3) \times
  A_4^\tree(-L_3, 3, 4, L_1)  \cr
&\hskip .3 cm 
 =  - i s t \, A_4^\tree(1, 2, 3, 4) \, 
   {1\over (L_1 - k_1)^2 } \, 
   {1\over (L_3 - k_3)^2 } \,, 
\cr} \hskip 2. cm 
$$
 it is a simple matter to evaluate
\eqn{GravitySewingStart}, yielding
$$
\hskip -.4 cm 
\eqalign{
& \sum_{N=8 \atop \rm\ states}  M_4^\tree(-L_1, 1, 2, L_3) \times
  M_4^\tree(-L_3, 3, 4, L_1) \cr \
& \hskip .2 cm
= i stu M_4^\tree(1, 2, 3, 4)
 \biggl[{1\over (L_1 - k_1)^2 } + {1\over (L_1 - k_2)^2} \biggr] \cr
& \hskip 3 cm \times
\biggl[{1\over (L_3 - k_3)^2 } + {1\over (L_3 - k_4)^2} \biggr]\,. \cr}
\equn\label{BasicGravCutting}
$$
The sewing equations for the $t$ and $u$ channels are similar to that
of the $s$ channel.

A remarkable feature of the cutting equation (\ref{BasicGravCutting})
is that the external-state dependence of the right-hand side is
entirely contained in the tree amplitude $M_4^{\rm \tree}$.  This fact
allows us to iterate the two-particle cut algebra to {\it all} loop
orders!  Although this is not sufficient to determine the
complete multi-loop four-point amplitudes, it does provide a wealth of
information.

Applying \eqn{BasicGravCutting} at one loop to each of the three
channels yields the one-loop four graviton amplitude of $N=8$
supergravity,
$$
\eqalign{
& {\cal M}_4^{\oneloop}(1, 2, 3, 4)
=  -i \Bigl( {\kappa \over 2}\Bigr)^4 
s t u  M_4^\tree(1,2,3,4)  \cr
& \hskip .2 cm \times
 \Bigl(  \I_4^{\oneloop}(s,t) 
           + \I_4^{\oneloop}(s,u)  
           + \I_4^{\oneloop}(t,u)  \Bigr) \,, \cr} 
$$
in agreement with previous results~\cite{GSB}.  We have reinserted the
gravitational coupling $\kappa$ in this expression.  The scalar
integrals are defined in \eqn{OneLoopIntegral} with ${\cal P} = 1$.

\begin{figure}[t]
\centerline{\psfig{figure=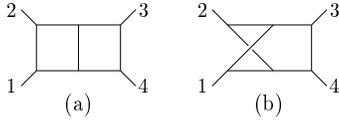,height=.7in}}
\vskip -.4cm 
\caption{The planar (a) and non-planar (b) scalar integrals, 
$\I_4^{\twoloop,\P}(s,t)$ and $ \I_4^{\twoloop,\NP}(s,t)$, appearing in
the two-loop $N=8$ amplitudes.  Each internal line
represents a scalar propagator.}
\label{fig:PlanarNonPlanar}
\end{figure}

At two loops, the two-particle cuts are given by a simple iteration of
the one-loop calculation. The three-particle cuts can be obtained by
recycling the corresponding cuts for the case of $N=4$
super-Yang-Mills.  It turns out that the three-particle cuts introduce
no other functions than those already detected in the two-particle cuts.
Combining all the cuts into a single function yields the $N=8$
supergravity two-loop amplitude~\cite{BDDPR},
$$
\hskip -.15 cm 
\eqalign{ 
& {\cal M}_4^{\twoloop}(1,2,3,4) = 
  \Bigl({\kappa \over 2} \Bigr)^6 \!\! stu  M_4^\tree(1,2,3,4) \cr
& \times
\Bigl(s^2 \, \I_4^{\twoloop,\P}(s,t) 
+ s^2 \, \I_4^{\twoloop,\P}(s,u)  \cr
& \hskip  .4 cm  
+ s^2 \, \I_4^{\twoloop,\NP}(s,t)
+ s^2 \, \I_4^{\twoloop,\NP}(s,u) \cr
& 
 \hskip 2  cm 
+\;  \hbox{cyclic} \Bigr) \,, \cr}
\equn\label{TwoLoopAmplitude}
$$
where `$+$~cyclic' instructs one to add the two cyclic permutations of
legs (2,3,4), and $\I_4^{\twoloop,\P/\NP}$ are depicted
in \fig{fig:PlanarNonPlanar}.

%%%%%%%%%
%%%%%%%

We comment that using the two-loop amplitude (\ref{TwoLoopAmplitude}),
Green, Kwon and Vanhove~\cite{GreenTwoLoop} provided an explicit
demonstration of the non-trivial M theory duality between $D=11$
supergravity and type II string theory.  Here the finite parts of the
amplitudes are most important, particularly their dependence on the
radii of one or two compactified dimensions.

%%%%%%%%%%%%%

\subsection{Divergence Properties of $N=8$ Supergravity}

Since the two-loop $N=8$ supergravity amplitude
(\ref{TwoLoopAmplitude}) has been expressed in terms of scalar
integrals, it is straightforward to extract the divergence properties.
The scalar integrals diverge only for $D\ge 7$; hence the two-loop
$N=8$ amplitude is manifestly finite in $D=5$ and $6$, contrary to
earlier expectations based on superspace power-counting
arguments~\cite{HoweStelle}. 
The discrepancy between the above explicit results and the earlier
superspace power counting arguments may be understood in terms of an
unaccounted higher dimensional gauge symmetry. Once this symmetry is
accounted for, superspace power counting gives the same degree of
divergence as the explicit calculation~\cite{StellePrivate}.

\iffalse

 The discrepancy between the above explicit
results and the previous superspace power counting arguments
may be understood as due to the lack of manifest $N=8$ supersymmetry in
the power counting. Once the symmetry is accounted for,
superspace power counting gives the same degree
of divergence as the explicit calculation~\cite{StellePrivate}.
\fi

The manifest $D$-independence of the cutting algebra allowed us to
extend the calculation to $D=11$, even though there is no
corresponding $D=11$ super-Yang-Mills theory. The result
(\ref{TwoLoopAmplitude}) then explicitly demonstrates that $N=1$ $D=11$
supergravity diverges.  In dimensional regularization there are no
one-loop divergences so the first potential divergence is at two
loops.  The $D=11$ two-loop divergence may be extracted from the
amplitude in \eqn{TwoLoopAmplitude} yielding~\cite{BDDPR},
$$
\eqalign{
&   {\cal M}_4^{\twoloop,\ D=11-2\e}\vert_{\rm pole} 
  = {1\over\e\ (4\pi)^{11}}
{1\over48} \, {\pi\over5\,791\,500} \cr
&  \hskip .3 cm 
\times  \Bigl( 438 (s^6+t^6+u^6) - 53 s^2 t^2 u^2 \Bigr) \cr
&  \hskip .3 cm
 \times \left( {\kappa \over 2}\right)^6 
                         \times stu M_4^{\rm tree} \, .} \hskip 1 cm 
$$
Further work on the structure of the $D=11$ counterterm has been
carried out in ref.~\cite{DeserSeminara}.

Since the two-particle cut sewing equation iterates to all loop
orders, one can compute all contributions which can be assembled
solely from two-particle cuts.  (The five-loop integral in
\fig{fig:Multiloop}, for example, falls into this category.)  Counting
powers of loop momenta in these contributions suggests the simple
finiteness formula, $L < 10/(D-2)$ (with $L>1$), where $L$ is the
number of loops.  This formula indicates that $N=8$ supergravity is
finite in some other cases where the previous superspace bounds
suggest divergences~\cite{HoweStelle}, e.g. $D=4$, $L=3$.  The first
$D=4$ counterterm detected via the two-particle cuts of four-point
amplitudes occurs at five, not three loops.  (A recent improved
superspace power count is, however, in agreement with this finiteness
formula~\cite{StellePrivate}.)  Further evidence that the finiteness
formula is correct stems from the MHV contributions to $m$-particle
cuts, in which the same supersymmetry cancellations occur as for the
two-particle cuts~\cite{BDDPR}.  However, further work would be
required to prove that other contributions do not alter the
two-particle cut power counting.  A related open question is whether one
can prove that the five-loop divergence encountered in the
two-particle cuts does not cancel against other contributions.

%%%%%%%%%%%%%%%%%%%%%%%%%%%%%%%%%%%%%%%%%%%

\section{Implications for the Einstein-Hilbert Lagrangian}

Consider the Einstein-Hilbert and Yang-Mills Lagrangians,
$$
L_{\rm EH} = {2 \over \kappa^2} \sqrt{-g} R \,, \hskip 1 cm 
L_{\rm YM} = - {1\over 4} F^a_{\mu\nu} F^{a\mu\nu}\,. \hskip 1 cm 
$$
The Einstein-Hilbert Lagrangian does not exhibit any obvious
factorization property that might explain the KLT relations.  We now
outline some recent work~\cite{Aaron} in interpreting the KLT
relations in terms of the Lagrangians.  (See also ref.~\cite{Siegel}.)

One of the key properties exhibited by the KLT relations
(\ref{KLTExamples}) is the separation of graviton Lorentz indices into
`left' and `right' sets.  Consider the graviton field, $h_{\mu\nu}$.
The KLT relations suggest that it should be possible to assign one
Lorentz index as being associated with a `left' gauge theory and the
other to a `right' one. Of course, since $h_{\mu\nu}$ is a symmetric
tensor it does not matter which index is assigned to the left or to
the right, but once the choice is made we would like to rearrange the
gravity Lagrangian so that left indices only contract with left ones
and right ones only with right ones.  Using the KLT relations we can
systematically find a Lagrangian with the desired properties
order-by-order in the perturbative expansion by reversing the usual
procedure of obtaining $S$-matrix elements from a Lagrangian.  In
ref.~\cite{Aaron}, an appropriate field redefinition for connecting
this Lagrangian to the standard Einstein-Hilbert one was described.

In conventional gauges, the difficulty of factorizing the
Einstein-Hilbert Lagrangian into left and right parts is already
apparent in the kinetic terms. In de Donder gauge, the quadratic part
of the Lagrangian is
$$
\hskip -2.2 cm 
L_2 = 
- \frac{1}{2} h_{\mu\nu} \partial^2 h_{\mu\nu} 
+ \frac{1}{4} h_{\mu\mu} \partial^2 h_{\nu\nu} \,,
\equn\label{QuadraticLagrang}
$$
so that the propagator is, 
$$
\hskip -.55 cm
\eqalign{
& P_{\mu\alpha;\nu\beta} = \cr
& {1 \over 2} 
\Bigl[\eta_{\mu\nu} \eta_{\alpha\beta} + \eta_{\mu\beta} \eta_{\nu\alpha} 
- {2\over D-2} \eta_{\mu\alpha} \eta_{\nu\beta} \Bigr] {i\over k^2 + i\eps}
\,. \cr}
$$
The appearance of the trace $h_{\mu\mu}$ in \eqn{QuadraticLagrang} is
unacceptable since it contracts a `left' and `right' index.  (In
Minkowski space, one of any two contracted indices should be raised
using $\eta^{\mu\nu}$, but this is suppressed here.)  Moreover, the
propagator contains explicit dependence on the dimension $D$, which
must somehow cancel from the tree-level $S$-matrix elements since
there is no such dependence in the KLT relations or in the gauge
theory amplitudes.

In order for the kinematic term to be consistent with the KLT
equations, all terms which contract a `left' Lorentz index with a
`right' one should be eliminated.  In ref.~\cite{BDS} a dilaton field
was introduced to allow for a field redefinition that removes the
graviton trace from the quadratic terms in the Lagrangian.  The
appearance of the dilaton as an auxiliary field to help rearrange the
Lagrangian is motivated by string theory which requires the presence
of such a field.  Following the discussion of refs.~\cite{BDS,Aaron},
consider the Lagrangian for gravity coupled to a dilaton,
$$
\hskip -2.25 cm 
L_{\rm EH} =   {2\over \kappa^2} \sqrt{-g} \, 
       R +  \sqrt{-g} \, \partial^\mu\phi\partial_\mu\phi \,.
$$
Since the auxiliary dilaton field is quadratic in the Lagrangian, it
does not appear in any tree diagrams involving only external
gravitons~\cite{Aaron}.  It therefore does not alter the tree
$S$-matrix of purely external gravitons.  (For theories containing
dilatons one can simply allow the dilaton to be an external physical
state.)  In de Donder gauge, for example, taking $g_{\mu\nu} =
\eta_{\mu\nu} + \kappa h_{\mu\nu}$, the quadratic part of the
Lagrangian including the dilaton is
$$
\hskip -1.15 cm 
L_2 = 
- \frac{1}{2} h_{\mu\nu} \partial^2 h_{\mu\nu} 
+ \frac{1}{4} h_{\mu\mu} \partial^2 h_{\nu\nu}
-  \phi \partial^2 \phi \,.
$$
The term involving $h_{\mu\mu}$ can be eliminated with the
field redefinitions,
$$
\hskip -3.05cm 
\eqalign{
& h_{\mu\nu} \rightarrow h_{\mu\nu} 
        + \eta_{\mu\nu}{\sqrt{\frac{2}{D-2}}}\, \phi\,, \cr
& \phi \rightarrow \frac{1}{2} h_{\mu\mu} + \sqrt{\frac{D-2}{2}}\, \phi\,,}
\equn\label{FieldRedef}
$$
yielding
$$
\hskip -3.1cm 
L_2 \rightarrow 
- \frac{1}{2} h_{\mu\nu} \partial^2 h_{\mu\nu} 
+  \phi \partial^2 \phi\,.
$$
One might be concerned that the field redefinition (\ref{FieldRedef})
would alter the gravity $S$-matrix.  However, the graviton $S$-matrix
is guaranteed to be invariant under non-linear field redefinitions or
under linear ones that do not alter the coupling to external traceless
tensors. 

Of course, the rearrangement of the quadratic terms is only the first
step.  In order to make the Einstein-Hilbert Lagrangian consistent
with the KLT factorization a set of field variables should exist where
all Lorentz indices can be separated into `left' and `right'
classes. To do so, all terms of the form
$$
\hskip -.01 cm 
h_{\mu\mu}\, , \hskip .5 cm 
h_{\mu\nu} h_{\nu\lambda} h_{\lambda\mu} \,, \hskip .5 cm 
\cdots, \hskip .5 cm \Tr[h^{2m+1}] \,,
\equn\label{BadTerms}
$$
where $\Tr[h^n] \equiv h_{\mu_1 \mu_2} h_{\mu_2 \mu_3} \cdots h_{\mu_n
\mu_1}$, need to be eliminated.  A field redefinition which
accomplishes this is,
$$
\eqalign{
g_{\mu\nu} & = e^{\sqrt{\frac{2}{D-2}} \kappa \phi} e^{\kappa\, h_{\mu\nu}} \cr
&  \equiv 
 e^{\sqrt{\frac{2}{D-2}} \,\kappa \phi} 
\Bigl(\eta_{\mu\nu} + \kappa h_{\mu\nu} + {\textstyle \kappa^2\over 2} 
                    h_{\mu\rho} h_{\rho\nu} + \cdots \Bigr) \,. \cr}
$$
As verified in ref.~\cite{Aaron} through $\Ord(h^6)$, this choice
eliminates all terms (\ref{BadTerms}) which mix left and right Lorentz
indices, even before gauge fixing.

It turns out that one can do better by performing further field
redefinitions and choosing a particular non-linear gauge.  With this
gauge it is possible to express the off-shell three graviton vertex in
terms of Yang-Mills three vertices~\cite{Aaron},
$$
\hskip -.2 cm 
\eqalign{
i & G_3^{\mu\alpha, \nu\beta, \rho\gamma}(k, p , q)\cr
 & = - {i \over 2} 
\Bigl( {\kappa \over 2} \Bigr)
\Bigl[
V_{\rm GN}^{\mu\nu\rho}(k, p, q) 
\times V_{\rm GN}^{\alpha \beta \gamma}(k, p, q) \cr
& \hskip 2 cm  
+V_{\rm GN}^{\nu\mu\rho}(p, k, q) \times 
V_{\rm GN}^{\beta \alpha \gamma}(p, k, q) \Bigr]\,, \cr }
$$
where
$$
\hskip -.3 cm 
 V^{\mu\nu\rho}_{\rm GN}(1,2,3) = 
       i\sqrt{2}  \bigl( k_1^\rho \, \eta^{\mu\nu}
            + k_2^\mu \eta^{\nu\rho} + k_3^\nu \eta^{\rho\mu} \bigr)
$$
is the color ordered Gervais-Neveu~\cite{GN} Yang-Mills three vertex,
from which the color factor has been stripped.  This is not the only
possible reorganization of the three vertex which respects the KLT
factorization.  The main reason for using the above vertex
is its simplicity.

This represents some initial steps in reorganizing the
Einstein-Hilbert Lagrangian so that it respects the KLT relations.  A
complete derivation of the KLT expressions starting from the
Einstein-Hilbert Lagrangian is, however, still lacking.

\section{Conclusions}

In this talk we described progress in understanding perturbative
quantum gravity thorough use of the Kawai-Lewellen-Tye string
relations~\cite{KLT} and $D$-dimensional
unitarity~\cite{SusyFour,Review}.  This provides an alternative way to
perturbatively quantize gravity theories, without direct 
reference to their Lagrangians or Hamiltonians, and leads to a relatively
efficient organization of the gravity $S$-matrix.  Known gauge theory
tree amplitudes can be recycled into gravity tree amplitudes which can
then be recycled into gravity loop amplitudes.

Using this approach, we described the argument of ref.~\cite{BDDPR}
that $D=4$ $N=8$ supergravity is less divergent than had been
previously believed.  Moreover, we outlined the computation of the
two-loop four-point amplitude in maximal supergravity given in
ref.~\cite{BDDPR}.  Using this result, it is straightforward to show
that $N=1$ $D=11$ supergravity diverges and is not protected by any
`magic' of M theory.  (Further elaboration of the structure of the
counterterm may be found in ref.~\cite{DeserSeminara}.)  The two-loop
supergravity amplitude (\ref{TwoLoopAmplitude}) has also been used by
Green, Kwon and Vanhove~\cite{GreenTwoLoop} to verify a non-trivial M
theory duality.

There are a number of interesting open questions. For example, the
methods described here have been used to investigate only maximal
supergravity. It would be interesting to systematically re-examine the
divergence structure of non-maximal theories.  (Some very recent work
on this may be found in ref.~\cite{DunbarJulia}.)  Using the 
methods described in this talk it might, for example, be possible to
systematically determine finiteness conditions order-by-order in the
loop expansion.  A direct derivation of the Kawai-Lewellen-Tye
decomposition of gravity amplitudes in terms of gauge theory ones
starting from the Einstein-Hilbert Lagrangian perhaps would lead to a
useful reformulation of gravity.  Connected with this is the question
of whether the heuristic notion that gravity is the square of gauge
theory can be given meaning outside of perturbation theory.  It
would also be interesting to know whether it is possible to relate more
general solutions of the classical equations of motion for gravity to
those for gauge theory.

In summary, in this talk we discussed how the heuristic notion
that gravity $\sim$ (gauge theory)$^2$ can be exploited to develop a
better understanding of perturbative gravity.

%%%%%%%%%%%%%%%%%%%%%%%%%%%%%%%%%%%%%%%%%%%%%%%%%%%%%%%%%

\end{document}